\documentclass[sigconf]{acmart}

\usepackage{xcolor}
\usepackage{listings}
\usepackage{color}
\usepackage{lipsum}
\usepackage[framemethod=TikZ]{mdframed}
\usepackage[ruled,vlined,linesnumbered]{algorithm2e}
\SetKwComment{Comment}{$\triangleright$\ }{}
\SetKwInOut{Variables}{Variables}

\usepackage{graphicx}
\usepackage{subfigure}
\usepackage{amsmath}
\usepackage{booktabs}
\usepackage{threeparttable}
\usepackage{multirow}
\usepackage{float}
\usepackage{makecell}
\usepackage{enumitem}
\usepackage{balance}

\newcommand{\eg}{\textit{e.g., }}
\newcommand{\ie}{\textit{i.e., }}
\newcommand{\etal}{\textit{et al. }}

\newcommand{\tool}{\textsc{Truzz}}

\definecolor{codegreen}{rgb}{0,0.6,0}
\definecolor{codegray}{rgb}{0.5,0.5,0.5}
\definecolor{codepurple}{rgb}{0.58,0,0.82}
\definecolor{backcolour}{rgb}{0.95,0.95,0.92}

\newcommand{\up}{\textcolor{codegreen}{$\uparrow$}}
\newcommand{\down}{\textcolor{red}{$\downarrow$}}

\DeclareMathOperator{\selectseed}{select\_seed}

\DeclareMathOperator{\fuzz}{execute}
\DeclareMathOperator{\sort}{sort}
\DeclareMathOperator{\update}{update}
\DeclareMathOperator{\mutate}{mutate}

\DeclareMathOperator{\pop}{pop}
\DeclareMathOperator{\mutatebytes}{mutate\_bytes}
\DeclareMathOperator{\getpath}{get\_path}
\DeclareMathOperator{\calculatefitness}{calculate\_fitness}
\DeclareMathOperator{\hasnext}{has\_next}
\DeclareMathOperator{\length}{length}
\DeclareMathOperator{\countmap}{count\_map}

\lstdefinestyle{mystyle}{
  backgroundcolor=\color{white}, commentstyle=\color{codegreen},
  keywordstyle=\color{magenta},
  numberstyle=\tiny\color{codegray},
  stringstyle=\color{codepurple},
  basicstyle=\ttfamily\footnotesize,
  breakatwhitespace=false,         
  breaklines=true,                 
  captionpos=b,     
  frame=single,
  keepspaces=true,                 
  numbers=left,                    
  numbersep=5pt,                  
  showspaces=false,                
  showstringspaces=false,
  showtabs=false,                  
  tabsize=2
}
\AtBeginDocument{%
  \providecommand\BibTeX{{%
    \normalfont B\kern-0.5em{\scshape i\kern-0.25em b}\kern-0.8em\TeX}}}

\copyrightyear{2022} 
\acmYear{2022} 
\setcopyright{acmlicensed}\acmConference[ICSE '22]{44th International Conference on Software Engineering}{May 21--29, 2022}{Pittsburgh, PA, USA}
\acmBooktitle{44th International Conference on Software Engineering (ICSE '22), May 21--29, 2022, Pittsburgh, PA, USA}
\acmPrice{15.00}
\acmDOI{10.1145/3510003.3510063}
\acmISBN{978-1-4503-9221-1/22/05}

\begin{document}

\title{Path Transitions Tell More: \\Optimizing Fuzzing Schedules via Runtime Program States}

\author{Kunpeng Zhang}
\affiliation{%
  \institution{Tsinghua Shenzhen International Graduate School, Tsinghua University}
  \country{} 
}
\email{zkp21@mails.tsinghua.edu.cn}

\author{Xi Xiao}
\affiliation{%
  \institution{Tsinghua Shenzhen International Graduate School, Tsinghua University}
  \country{} 
}
\email{xiaox@sz.tsinghua.edu.cn}

\author{Xiaogang Zhu{$^\ast$}}
\affiliation{%
  \institution{Swinburne University of Technology}
  \country{} 
}
\email{xiaogangzhu@swin.edu.au}

\author{Ruoxi Sun}
\affiliation{%
  \institution{The University of Adelaide}
  \country{} 
}
\email{ruoxi.sun@adelaide.edu.au}

\author{Minhui Xue}
\affiliation{%
  \institution{The University of Adelaide}
  \country{}
}
\email{jason.xue@adelaide.edu.au}

\author{Sheng Wen{$^\ast$}}
\affiliation{%
  \institution{Swinburne University of Technology}
  \country{}
}
\email{swen@swin.edu.au}

\renewcommand{\shortauthors}{Kunpeng Zhang, et al.}
\renewcommand \authors{Kunpeng Zhang, Xi Xiao, Xiaogang Zhu, Ruoxi Sun, Minhui Xue, and Sheng Wen}

\begin{abstract}
Coverage-guided Greybox Fuzzing (CGF) is one of the most successful and widely-used techniques for bug hunting. Two major approaches are adopted to optimize CGF: \textit{(i)} to reduce search space of inputs by inferring relationships between input bytes and path constraints; \textit{(ii)} to formulate fuzzing processes (\eg path transitions) and build up probability distributions to optimize power schedules, \ie the number of inputs generated per seed. However, the former is subjective to the inference results which may include extra bytes for a path constraint, thereby limiting the efficiency of path constraints resolution, code coverage discovery, and bugs exposure; the latter formalization, concentrating on power schedules for seeds alone, is inattentive to the schedule for bytes in a seed. 

In this paper, we propose a lightweight fuzzing approach, \tool{}, to optimize existing Coverage-guided Greybox Fuzzers (CGFs). To address two aforementioned challenges, \tool{} identifies the bytes related to the validation checks (\ie the checks guarding error-handling code), and protects those bytes from being frequently mutated, making most generated inputs examine the functionalities of programs, in lieu of being rejected by validation checks. The byte-wise relationship determination mitigates the problem of loading extra bytes when fuzzers infer the byte-constraint relation. 
Furthermore, the proposed path transition within \tool{} can efficiently prioritize the seed as the new path, harvesting many new edges, and the new path likely belongs to a code region with many undiscovered code lines. 
To evaluate our approach, we implemented 6 state-of-the-art fuzzers, AFL, AFLFast, NEUZZ, MOPT, FuzzFactory and GreyOne, in \tool{}. 
The experimental results show that on average, \tool{} can generate 16.14\% more inputs flowing into functional code, in addition to 24.75\% more new edges than the vanilla fuzzers.
Finally, our approach exposes 13 bugs in 8 target programs, and 6 of  them have not been identified by the vanilla fuzzers. {\let\thefootnote\relax\footnote{{$\ast$~Corresponding authors: Sheng Wen and Xiaogang Zhu}}}
\end{abstract}




\maketitle

\section{Introduction}
Fuzzing is a widely-used technique to expose security issues in software, and has identified thousands of bugs and vulnerabilities in real-world programs \cite{ossfuzz, onefuzz, feng2021snipuzz}. Coverage-guided Greybox Fuzzing (CGF) is one of the most successful solutions in fuzzing \cite{afl, redqueen, steelix, vuzzer, GreyOne, aflfast, Yue2020EcoFuzz}. CGF usually maintains an infinite loop, where new inputs are generated by mutating seeds. If a new input explores new code coverage (\eg new edges), the input will be retained as a new seed and mutated in further rounds. This evolutionary solution effectively guides fuzzing towards exploring more code coverage.

Existing solutions to improve the efficiency of CGF are two-fold. The first category intends to reduce search space of inputs~\cite{angora,redqueen,steelix, neuzz, vuzzer, GreyOne}. 
The state-of-the-art solution to reduce the input space is to infer the relationships between the input bytes and the path constraints~\cite{neuzz, GreyOne}, \ie the byte-constraint relation, which significantly improves the possibility of resolving path constraints. 
For example, an input includes 8 bytes but only 1 byte is related to a constraint,  \texttt{if(x[5]==0xee)}. If CGFs can infer the byte-constraint relation, the mutation could be focused on the related byte, \texttt{x[5]}, only.
As a result, instead of trying $256^8$ new inputs, CGFs only need to generate at most $256$ new inputs to resolve the constraint. 
However, a real-world program could be far more complicated as the byte-constraint relation could exist between multiple bytes and multiple different constraints, or even nested. 
From our experience, the existing inference approaches cannot distinguish between the outer condition and the inner condition of a nested one.
For example, the inferred bytes are likely related to an entire nested condition (\eg \texttt{if(x[5]==0xee) \{if(x[6]==0xff) \{...\} \}}). 
That is, when the inference process intends to infer the related bytes for the inner condition (\ie \texttt{x[6]}), the result will erroneously include the bytes related to the entire condition (\ie \texttt{x[5],x[6]}).
As a result, when fuzzing intends to resolve the inner condition, fuzzing may mutate the byte \texttt{x[5]}, which will alter the branch of outer condition. 
Thus, the generated input will never reach the inner condition \texttt{if(x[6]==0xff)} because the mutation of \texttt{x[5]} breaches the outer condition.

On the other hand, another category of solutions optimizes power schedule via formulating fuzzing processes \cite{aflfast, Boehme2017Directed, Yue2020EcoFuzz, bohme2020boosting, mtfuzz}.
Because CGFs aim to discover more code coverage, one state-of-the-art solution to optimize power schedule is to model the path transitions as Markov chain~\cite{aflfast} or Multi-Armed Bandit~\cite{Yue2020EcoFuzz}. The \textit{power schedule} determines the number of mutations assigned to each seed. A new input  $s'$, which exercises the path $P_{s'}$, is generated by mutating a seed $s$, which exercises the path $P_s$. Then, there is a path transition between the path $P_s$ and $P_{s'}$. The models build probability distributions of path transitions, and optimize power schedule based on the distributions. A basic motivation is to assign more power to seeds that are more likely to discover new coverage, \ie they prefer the path transitions including new paths. For example, based on Markov chain, AFLFast~\cite{aflfast} prefers seeds that exercise less frequent paths, \ie the seeds with lower transition probabilities. However, the existing models only focus on power schedule for seeds but ignore the power schedule for bytes. In fact, the transitions contain more information of the execution states, which can be utilized to optimize power schedule for bytes.

\begin{figure}[t]
\centering
\includegraphics[width=\linewidth]{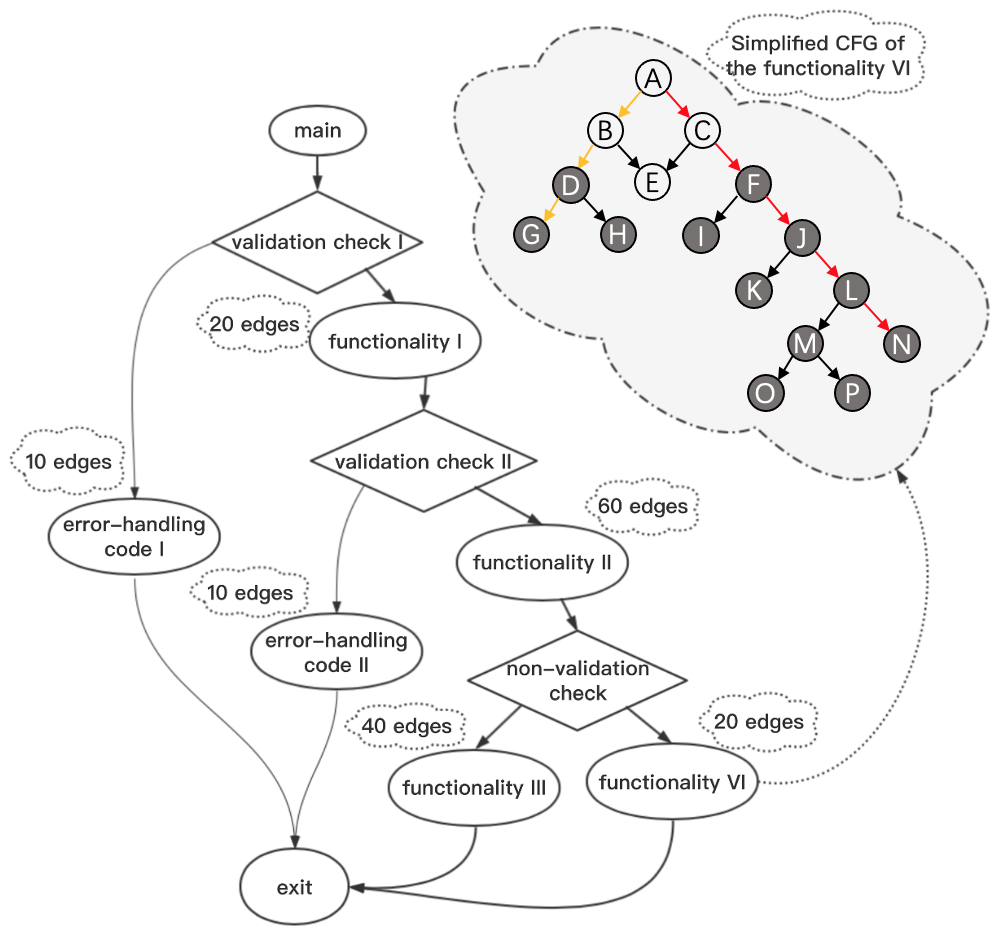}
\caption{An example of validation checks and code coverage. The execution paths containing error-handling code is likely to be shorter than the paths of functionalities. As for code coverage, a newly discovered path that includes more new edges is likely to have more undiscovered descendants.}
\label{fig:validation_check}
\end{figure}

We observe that the differences between two execution paths of a transition imply properties of programs under test (PUTs). Specifically, if a path transition results in a significantly shorter path, the transition is likely related to a validation check. 
In this research, we follow the definition in You \etal~\cite{profuzzer}: if a path constraint guards error-handling code, then it is a validation check; otherwise, it is a non-validation check. 
The execution paths that examine the error-handling code are usually much shorter than the paths that examine functional code. The reason lies in the fact that the error-handling code usually leads to the termination of a program. As shown in Fig.~\ref{fig:validation_check}, if an input fails the \textit{validation check I}, the input will flow into the \textit{error-handling code I}, whose length is 10 edges. On the other hand, if another input passes the \textit{validation check I} and \textit{validation check II}, the length of the execution path is at least 100 edges.

Moreover, if a path transition discovers an execution path that includes new edges, it is likely to explore an undiscovered region.  
Intuitively, more new edges covered in a path may lead us to more undiscovered descendants as it may have touched more new code areas, so that we should focus more on such seeds that discovered more new edges until they have been thoroughly tested. 
Considering the limited time budget of fuzzing campaigns, the prioritization of seeds that exercise paths with more undiscovered descendants can further improve the efficiency of code discovery. 
We show in Fig.~\ref{fig:validation_check} that if the new edges in the path $ACFJLN$ are $CF$, $FJ$, $JL$, and $LN$, the undiscovered basic blocks that the path $ACFJLN$ can reach are $I$, $K$, $M$, $O$, and $P$; on the other hand, if another path $ABDG$ includes new edges $BD$ and $DG$, the undiscovered basic block that the path $ABDG$ can reach is $H$. Therefore, the input that exercises the new path $ACFJLN$ is more likely to explore larger undiscovered code regions. 

In this paper, we propose a lightweight approach, \textbf{\tool{}}, to improve the performance of CGFs. Different from other existing CGFs, our \tool{} determines the validation-related bytes without complex static analysis and dynamic analysis. Concretely, \tool{} intends to protect bytes related to validation checks, and prioritizes seeds that are more likely to explore larger new code regions. It infers the relations between input bytes and validation checks. Based on the aforementioned observation, if mutation of certain bytes results in exercising a significantly shorter path, then the bytes are likely related to a validation check. In order to determine such a relationship, \tool{} computes a scalar, the \textit{fitness}, for each byte, measuring how likely a byte is related to a validation check. 
Based on the fitness score, \tool{} then assigns lower mutation probabilities to the validation-related bytes, locking them from being mutated. 
Furthermore, in order to prioritize seeds, \tool{} ranks each seed based on the number of newly discovered edges when a seed is first retained. 
The top seed will be selected to be mutated in the next fuzzing round.

The experimental results indicate that the integration of our approach and the existing CGFs can further improve the performance of CGFs. 
Our approach can help mitigate the aforementioned problem of the imprecise inference of byte-constraint relations. 
It focuses on the fine granularity of byte schedule, which can improve the performance of coverage discovery.
Meanwhile, our approach steers computing resources towards functional code rather than error-handling code. 
Due to the limited time budget, the Seed Prioritization in our approach further improves the performance of power schedule.

We evaluate the performance of \tool{} by applying it onto 6 state-of-the-art fuzzers, \ie NEUZZ~\cite{neuzz}, GreyOne~\cite{GreyOne}, AFL~\cite{afl}, MOPT~\cite{MOPT}, FuzzFactory~\cite{fuzzfactory}, and AFLFast~\cite{aflfast}, and conduct fuzz testing on 8 target programs. 
The experimental results show that \textsc{Truzz}-equipped fuzzers discover more code coverage and bugs than the vanilla fuzzers. 
We summarize our contributions as follows.

\begin{itemize}[leftmargin=*]
    \item We \textit{first} propose a lightweight generalized framework of byte schedule and Seed Prioritization, \tool, to improve the performance of CGFs, utilizing the differences of paths in path transitions, as such differences can indicate properties of target programs, \eg the validation checks and the undiscovered code regions. 
    \item We evaluate our approach with respect to valid input generation. The experimental results indicate that, with the help of \tool{}, 16.14\% on average more inputs can pass the validation checks and flow into functional code. 
    \item We implement our framework with 6 state-of-the-art fuzzers, AFL, AFLFast, NEUZZ, MOPT, FuzzFactory, and GreyOne, and compare the performance with vanilla fuzzers. On average 24.75\% more edge coverage has been identified. Moreover, \textsc{Truzz}-equipped fuzzers expose 13 bugs in 8 target programs, and 6 of them have not been identified by the vanilla fuzzers.
\end{itemize}

\section{Related Work and Motivation}
In this section, we introduce the related work of Coverage-guided Greybox Fuzzing (CGF). We then attempt to motivate our approach by demonstrating the weaknesses of current CGFs.  

\subsection{Coverage-guided Greybox Fuzzing} \label{sec:fuzzing_background}

The basic idea of CGF is to explore more code coverage so that it may trigger a bug hiding in certain code regions~\cite{fairfuzz, aflsmart, collafl, superion, nautilus, learnafl, Zhu2020CSI}. 
A general CGF has a seed corpus which contains the initial inputs and ``interesting'' inputs that are retained from the generated inputs during fuzzing. The initial inputs are usually manually created according to the input specifications of PUTs or collected from the Internet. The ``interesting'' inputs are the ones that try to discover new code coverage during fuzzing. Specifically, during each round of fuzzing, CGF selects a seed from the seed corpus and generates new inputs by performing mutations on the bytes in the seed~\cite{honggfuzz, cha2015program, duran2011targeted}. Most CGFs randomly select byte(s) to be mutated~\cite{afl,aflfast} or select byte(s) that are related to a certain path constraint~\cite{angora, GreyOne, neuzz}. When a generated input discovers new code coverage (\eg new edges), CGF will retain the input as a new seed. The loop that includes selection, mutation, and retention ensures the effectiveness of code coverage discovery~\cite{afl, aflfast, aflchurn}.

Most CGFs obtain the coverage information via instrumenting a bitmap to indicate new edges. A bitmap is a compact vector that records the discovered edges in the current execution. Each edge is assigned with an identifier which is also the index of the bitmap. When an edge is examined, the corresponding element of the bitmap will increase the value by one. As a result, if the value of an element is 0, the corresponding edge is not examined; otherwise, if the value of an element is non-zero, the corresponding edge is examined. Therefore, a new edge is determined if the corresponding element of the bitmap has flipped from 0 to non-zero in the current execution.

\subsubsection{Path Constraints}
A major roadrock of fuzzing efficiency improvement is the path constraints, especially the tight constraints, such as \texttt{if(x[5]==0xee)}. Many fuzzers utilize symbolic execution to resolve path constraints, but suffer from significantly slow execution speed~\cite{driller, qsym}. Because the bytes related to a path constraint are usually a small portion of an input, some other fuzzers utilize taint analysis to build the relationships between input bytes and path constraints~\cite{taintscope,vuzzer, steelix, angora,tiff}. Taint tracking can identify promising input bytes that affect program's certain operations~\cite{argos,whiteboxtaint,dowsing,taintdroid,taintart,neutaint}. When resolving a path constraint, fuzzing only needs to mutate the related bytes so that they improve the efficiency of passing raodrocks. However, taint analysis still introduces high overhead that limits the improvement of performance. T-Fuzz~\cite{tfuzz} utilizes program transformation to remove path checks so that fuzzing can explore deep code, but it requires many engineering efforts to transform programs.

\begin{figure}[t!]
\centering
\includegraphics[width=0.9\linewidth]{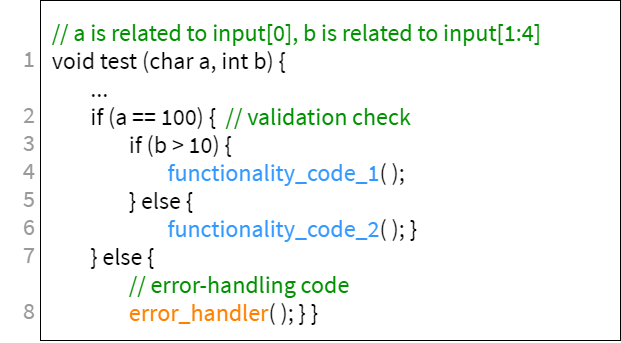}
\caption{Validation check. If an input fails the validation check, it will be trapped in the error-handling code.}
\label{lst:motivate_example}
\end{figure}

More lightweight approaches are proposed to resolve path constraints, which infer the relationships between input bytes and path constraints based on the changes of execution states~\cite{GreyOne, neuzz,mtfuzz,redqueen}. For instance, NEUZZ~\cite{neuzz} builds up deep learning models for the relationships between input bytes and branch behaviors. The branch behaviors indicate the execution states of satisfied or unsatisfied path constraints. The models have a high chance to include extra bytes for a certain path constraint because they approximately build the relationships~\cite{neuzz}.
Similarly, the inference approach of GreyOne~\cite{GreyOne} is that, if mutation of a byte results in changing the value of a variable, then the byte is related to the variable. When the variable is checked in a path constraint, the byte is also related to the path constraint. However, the change of bytes related to the outer conditions can also influence the value of variables related to the inner conditions (see Fig.~\ref{lst:motivate_example}). Therefore, the related bytes for inner conditions will include extra bytes. 
Due to extra bytes, fuzzing may mutate the bytes related to the outer condition, which will alter the branch of outer condition, resulting in decreasing the effectiveness of resolving the path constraint of inner condition. 

\subsubsection{Power Schedule Optimization}
Another way to improve the performance of CGF is to optimize fuzzing schedules via formulating fuzzing processes~\cite{MOPT,aflfast,Yue2020EcoFuzz,bohme2020boosting}. 
For general CGFs (\eg AFL~\cite{afl}), they assign almost the same energy 
for each seed. However, different seeds have different potentials to discover new coverage. Therefore, based on formulating fuzzing processes, CGFs~\cite{Yue2020EcoFuzz, aflfast} assign more energy to seeds that are more likely to discover new coverage. A 
optimization for power schedule is based on the formulation of path transitions. 
CGFs build the probability distributions of path transitions via observing the result of each mutation. 
More energy will be assigned to the seeds that have higher probabilities to transition to an undiscovered path. 
However, they only focus on the power schedule for seeds but ignore the schedule for bytes.

\begin{figure*}[t!]
\centering 
\includegraphics[width=0.9\linewidth]{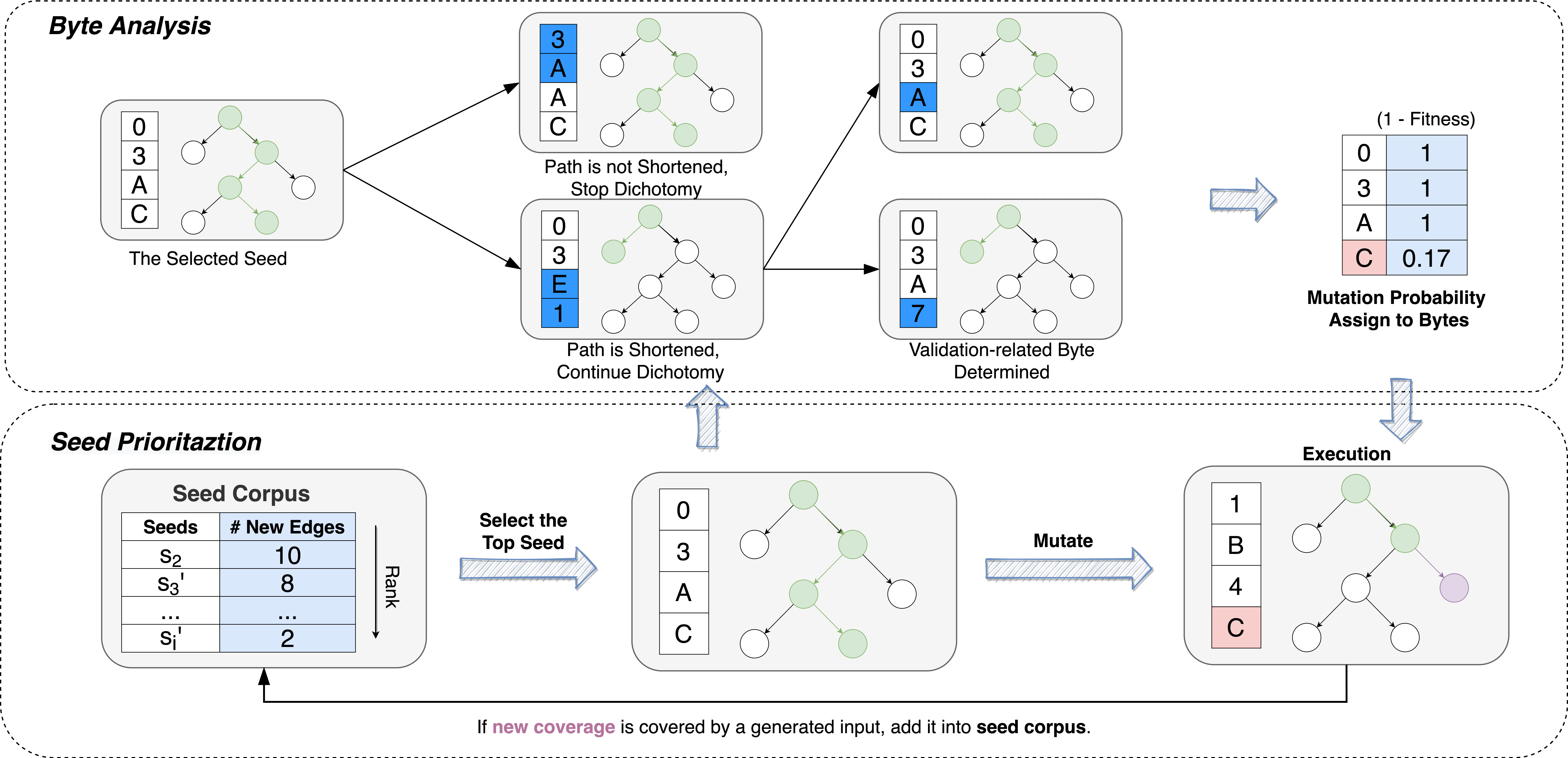}
\caption{Overview of \tool. \tool{} ranks each seed and prioritizes seeds with higher ranks. When a seed is selected, \tool{} will perform \textit{Byte Analysis} to identify the bytes related to validation checks. Then, during the mutation stage, it will avoid mutating the protected bytes with a high probability.}
\label{fig:overview}
\end{figure*}

\subsection{Motivating Example}
\tool{} improves the performance of CGFs based on the differences between path transitions. 
In this section, we show an example that motivates this research, \ie the requirement for the byte schedule.

As shown in Fig.~\ref{lst:motivate_example}, line 2 is a validation check because it guards the error-handling code in line 8. In order to test the functional code (lines 4 and 6), fuzzing has to satisfy the validation check in line 2. On the other hand, if a seed has already satisfied the condition in line 2, the later fuzzing campaigns should not mutate the bytes related to the condition. Therefore, some state-of-the-art fuzzers~\cite{neuzz, GreyOne, mtfuzz} improve the efficiency of resolving the path constraints via inferring the byte-constraint relation. For example, the length of input in Fig.~\ref{lst:motivate_example} is 5 bytes and a random mutation may generate $256^5$ new inputs and most of them will be rejected by the validation check in line 2. 
However, if a fuzzer can infer that the first byte is the only byte related to the constraint, it only needs at most $256$ mutations to bypass the constraint.

The problem is that the inference may include extra bytes for a path constraint. 
The reason lies in the fact that the inference relies on the changes of execution states, and the changes may be explicitly or implicitly influenced by certain bytes.
For example, the constraint of line 3 in Fig.~\ref{lst:motivate_example} is explicitly influenced by  variable $b$ and implicitly influenced by variable $a$. Therefore, for both NEUZZ and GreyOne, the bytes related to line 3 are $input[0:4]$, \ie the bytes related to both $a$ and $b$. As a result, fuzzers may mutate the byte $input[0]$ when resolving the constraint of line 3, and generate many inputs that examine the code \textit{error\_handler()} rather than the functionalities. Therefore, in order to examine the functionalities, the byte $input[0]$ that satisfies the check should not be mutated.

Our \tool{} mitigates the aforementioned problem via identifying the bytes related to validation checks based on path transition. When \tool{} mutates the related bytes, it will protect the bytes related to the validation checks (\eg the byte $input[0]$ in Fig.~\ref{lst:motivate_example}). Therefore, most of the generated inputs examine the functional code and avoid being trapped in the error-handling code. Meanwhile, the path transitions are also utilized to prioritize seeds, which can further improve the efficiency of CGFs.

\section{Methodology of \tool{}}

Motivated by the aforementioned observation that the differences in path transitions imply internal states of programs, we design the \tool{} as a lightweight framework to improve the code coverage and efficiency of CGFs. Our \tool{} is lightweight because it does not utilize complex taint analysis~\cite{angora, steelix} to obtain related bytes or approaches to pinpoint the error-handling code regions~\cite{FIFUZZ}. As shown in Fig.~\ref{fig:overview}, \tool{} consists of two core components, Byte Analysis and Seed Prioritization. 

\tool{} performs Byte Analysis on each seed to identify the bytes related to validation checks. It determines the validation-related bytes if mutation of those bytes results in significantly shorter execution paths. Since validation-related bytes are sparse in inputs, \tool{} utilizes dichotomy to improve the efficiency of the process.  
Note that the Byte Analysis is only performed once for each seed, which introduces small overhead. During the mutation stage, the validation-related bytes are protected so that most of the generated inputs examine the functional code, instead of being rejected by the validation checks. 
Then, in Seed Prioritization, \tool{} ranks each seed when they are added to the seed corpus. The seeds with more newly discovered edges are tagged with higher rankings, which are utilized to prioritize seeds. The combination of Byte Analysis and Seed Prioritization intends to optimize power schedule in fuzzing via the balance of “\textit{exploitation} vs. \textit{exploration}”. Byte Analysis focuses more on the “\textit{exploitation}” because it puts more efforts on fuzzing functionality code. On the other hand, Seed Prioritization intends to “\textit{explore}” because it prefers seeds that have more potential to discover more code regions.

\subsection{Byte Analysis}
In this stage, \tool{} builds up a probability distribution for input bytes, and selects bytes to be mutated based on the distribution. The probabilities are gained based on the differences between two execution paths in a path transition. Essentially, if a byte is more likely related to validation checks, the byte will be selected with a smaller probability. As a result, the validation-related bytes are protected during fuzzing, and the generated inputs are more likely to explore the functionalities of programs. 
\tool{} first measures how likely a byte is related to a validation check as a scalar, the \textit{fitness}, and further assigns a mutation probability to the byte.

\subsubsection{Fitness for Bytes}
In order to measure how likely a byte is related to validation checks, \tool{} first mutates each byte and computes the \textit{fitness} based on the differences in the path transition. 
Equation~\ref{equ:fitness} defines the fitness $f(i,m)$ for the $i^{th}$ to $(i+m)^{th}$ bytes. Specifically, let $s$ be a seed, and $s'$ be the generated input obtained by mutating the $i^{th}$ to $(i+m)^{th}$ bytes of the seed. 
Suppose that the seed~$s$ exercises the path $P_{s}$ and the input $s'$ exercises the path~$P_{s'}$. Note that, a path consists of a set of edges in this context. In the equation, $P_{s} \cap P_{s'}$ is the subset of edges that belong to both paths~$P_{s}$ and $P_{s'}$, and $|\ast|$ presents the number of edges in a set. For example, $|P_s \cap P_{s'}|$ indicates the number of edges that belong to the intersection of paths $P_{s}$ and $P_{s'}$.

Recall that, if an input exercises an execution path that contain error-handling code, the length (\ie the number of edges) may be significantly shorter than the path exercised by the corresponding seed. In Equation~\ref{equ:fitness}, $\frac{|P_s|-|P_{s'}|}{|P_s|}$ indicates how significantly the length of path has been shortened. 
Further, $\frac{|P_s \cap P_{s'}|}{|P_s|}$ measures the intersection of the execution paths $P_{s}$ and $P_{s'}$, \ie the edges changed in the path transition. 
The reason we not only consider the length of paths is that, when the difference of length and the intersection of paths are both small, the two paths will be quite different but have similar lengths. In such (rare) cases, the bytes having been mutated are more likely to be related to a validation check and the \textit{fitness} should be large.
Therefore, Equation~\ref{equ:fitness} measures the differences in a path transition pertaining to the changes of both length and individual edges. 
\begin{small}
\begin{equation} \label{equ:fitness}
f(i,m) = \begin{cases}
$$\begin{split}
&\frac{1}{2} \times [\frac{|P_s|-|P_{s'}|}{|P_s|} + (1-\frac{|P_s \cap P_{s'}|}{|P_s|})] \\
&=1-\frac{|P_{s'}|+|P_s \cap P_{s'}|}{2|P_s|}
\end{split}$$, & |P_s|>|P_{s'}| \\
0, & |P_s| \leq |P_{s'}| \\
\end{cases}
\end{equation}
\end{small}

Here we take Fig.~\ref{fig:validation_check} as an example for computing \textit{fitness} $f(i,m)$.
Suppose that a valid seed examines the \textit{functional code III} and explores $P_s = 20 + 60 + 40 = 120$ edges. If mutating the $a^{th}$ byte of the seed generates an input that flows into \textit{error-handling code II}, the input explores $P_{s'} = 20 + 10 = 30$ edges. Based on Equation~\ref{equ:fitness}, the fitness is calculated as $f(a,0) = \frac{1}{2} \times [\frac{120-30}{120} + (1-\frac{20}{120})] \approx 0.79 $. If we mutate the $b^{th}$ to $(b+c)^{th}$ bytes and the generated input examines the \textit{functional code VI}, the new input explores $20 + 60 + 20 = 100$ edges. Then, the fitness is $f(b,c) = \frac{1}{2} \times [\frac{120-100}{120} + (1-\frac{80}{120})] = 0.25$. 
Thus, \tool{} infers that the $a^{th}$ byte is more likely to be related to validation checks than the $b^{th}$ to $(b+c)^{th}$ bytes.

\subsubsection{Dichotomy for Byte Analysis}

The calculation of fitness byte by byte is time-consuming, especially when an input includes a large number of bytes. Fortunately, validation checks are sparse in programs, \ie only a small portion of input bytes are related to validation checks. Most of the input bytes flow into the code of functionalities. With this in mind, we utilize a dichotomy method to improve the computing efficiency. If mutation of the bytes belonging to the current segment results in a fitness lower than the threshold or the size of the segment is small enough, all bytes in the segment are set to the same fitness; otherwise, \tool{} continues the dichotomy until the validation-related bytes are identified.

As shown in Algorithm~\ref{alg:byte_analysis},
\tool{} first creates an array $F$ presenting the \textit{fitness} value for each byte (line 1) and initializes a set $D$ with the left and right half intervals of a seed (lines 2-4). Then the dichotomy starts. During the loop, if $D$ is not empty, \tool{} obtains an interval denoted by the starting and ending indexes, mutates the bytes, and determines the path $P_{s'}$, which is covered by the mutated input $s'$ via the bitmap $B'$ (lines 5-9). With the paths $P_{s}$ and $P_{s'}$, according to Equation~\ref{equ:fitness}, \tool{} computes the \textit{fitness} for bytes in the current interval (line 10). If the $fitness$ is smaller than a predefined threshold or the length of the current interval is less than the predefined minimum length, all the bytes in the interval are assigned with the value of $fitness$ (lines 11-12). Otherwise, \tool{} continues the dichotomy by creating new intervals and appends them into the set $D$ (lines 14-15). Considering that in each round of dichotomy, \tool{} is able to  handle at most half of the bytes within one mutated input, so the assignment of fitness for bytes performs in an efficient manner.

For example, as shown in Fig.~\ref{fig:overview}, the seed for the target program is  \textit{\underline{03AC}}. During the stage of Byte Analysis, \tool{} mutates the first two bytes \textit{\underline{03}}, and generates the input \textit{\underline{3AAC}}. Because the execution paths exercised by \textit{\underline{03AC}} and \textit{\underline{3AAC}} do not change, \tool{} assigns fitness to the first two bytes as 0, which is definitely smaller than the threshold. On the contrary, when mutating the other half bytes \textit{\underline{AC}}, the execution path is significantly shorter and the \textit{fitness} value is 0.83, which is larger than the threshold, so that the dichotomy continues. Then, \tool{} continues to split the bytes \textit{\underline{AC}} into two halves \textit{\underline{A}} and \textit{\underline{C}}. Finally, \tool{} infers that the last byte \textit{\underline{C}} is related to the validation check. Based on dichotomy, \tool{} only needs to conduct roughly $2\times log_2(N)$ mutations, where $N$ is the length of an input. Note that a larger threshold of fitness and a larger threshold of interval size will reduce the computing cost but lead to a coarse-grained result.

\begin{algorithm}[!t]
\footnotesize
\caption{Dichotomy for Byte Analysis}\label{alg:byte_analysis}
\KwIn{The current seed, $s$. 
The path exercised by $s$, $P_s$.
The minimum length of interval, $l$.
The threshold of fitness, $t$.}
\Variables{
A Bitmap that determines code coverage, $B$.
The start index of an interval, $b$.
The end index of an interval, $e$.
A set of $(b, e)$ intervals, $D$.
The fitness, $f$.
}
\KwOut{An array of $f$ for each byte, $F$.
}
$size \gets \length(s) - 1$\;
$F[0:size] \gets 0$\;
$D \gets D \cup (0, size/2)$\;
$D \gets D \cup (size/2 + 1, size)$\; \label{line:init_half}
\While{$\hasnext(D)$}{
    $(b,e) \gets \pop(D)$\; \label{line:next_interval}
    $s' \gets \mutatebytes(s, b, e)$\;
    $B' \gets \fuzz(s')$\;
    $P_{s'} \gets \getpath(B')$\;
    $f \gets \calculatefitness(P_{s}, P_{s'})$\; \label{line:cal_fitness}
    
    \eIf{$f < t\ or\ e - b < l$}{
        $F[b:e] \gets f$\;\label{line:record}
    } 
    {
        $D \gets D \cup (b, (b+e)/2)$ \;\label{line:lhalf}
        $D \gets D \cup ((b+e)/2+1, e)$ \;\label{line:append}
    }
}
\end{algorithm}

\subsubsection{Probability for Bytes Mutation}
To protect bytes related to validation checks, based on the \textit{fitness}, \tool{} decides whether a byte can be mutated or not when generating a new input.
The purpose of assigning a mutation probability to each byte is to avoid mutating the validation-related bytes with a high probability. 
\tool{} assigns the mutation probability according to Equation~\ref{equ:prob}.

\begin{equation}\label{equ:prob}
    p(i,m) = max(1 - f(i,m), L_p), 
\end{equation}
where $f(i,m)$ is the \textit{fitness} of $i^{th}$ to $(i+m)^{th}$ bytes. A larger fitness of a byte will lead to a smaller probability to be mutated, so that a validation-related byte could be protected from mutation. 
Note that we also set a lower bound of the probability $L_p$ to ensure that fuzzing will still generate inputs to fail the validation checks and cover the error-handling code, instead of ignoring it at all.

\subsection{Seed Prioritization}

\begin{algorithm}[!t]
\footnotesize
\caption{Seed Prioritization}\label{alg:seed_prioritization}
\KwIn{Seed Corpus, $S$. The number of mutations, $energy$.}
\Variables{The number of new edges, $n$. The number of new edges found by mutating one seed, $n_{all}$. 
A set of (seed, \# new\_edges) tuples, $M$.
A generated input, $i$. A bitmap that tracks code coverage, $B$.}
$B_{overall} \gets \emptyset$\;
\For{$s \in S$}{
    $B \gets \fuzz(s)$\;
    $n \gets \countmap(B \setminus (B_{overall} \cap B))$\;
    \If{$n > 0$}{
            $M \gets M \cup (s,n)$ \;
        }
    $B_{overall} \gets B_{overall} \cup B$\;
}
$M \gets \sort(M)$\;
\While{$True$}{
    $s \gets \selectseed(M)$\;
    $n_{all} \gets 0$\;
    \While{$energy > 0$}{
        $i \gets \mutate(s)$ \Comment*[r]{Based on Byte Analysis}
        $B' \gets \fuzz(i)$\;
        $n' \gets \countmap(B' \setminus (B_{overall} \cap B'))$\;
        \If{$n' > 0$}{
            $M \gets M \cup (i,n')$ \;
        }
        $energy \gets energy - 1$\;
        $B_{overall} \gets B_{overall} \cup B'$\;
        $n_{all} \gets n_{all} + n'$\;
    }
    $M \gets \update (s,n_{all})$\;
    $M \gets \sort(M)$\;
}
\end{algorithm}

Most CGFs sequentially select seeds to be mutated~\cite{afl, redqueen, steelix, vuzzer, aflfast, mtfuzz, neuzz}, \ie they select seeds based on the order of the seeds added to the seed corpus. 
However, seeds differ from their potential to explore undiscovered code regions. Due to the limited time budget, this scheme may miss opportunities to discover more code coverage. In order to improve the efficiency of code discovery, \tool{} heuristically determines the potential for each seed to discover more code coverage. Intuitively, if a newly discovered path includes more new edges, the path is likely to explore a code region with more undiscovered code lines. That is, a new path that includes more new edges has a higher chance to have more descendants of undiscovered blocks (\eg the blocks $I,K,M,O,P$ in Fig.~\ref{fig:validation_check}). Therefore, \tool{} ranks each seed based on the number of new edges in a newly discovered execution path. 
The larger the number of new edges, the higher priority to select the corresponding seed. 
Note that, \tool{} updates the rankings of each seed during fuzzing campaigns. If a seed discovers only a few new edges after being fuzzed many times, \tool{} decreases the ranking of the seed.

As shown in Algorithm~\ref{alg:seed_prioritization}, \tool{} first performs a dry run to assign rankings for initial seeds (lines 1-8). 
The bitmap $B_{overall}$ tracks all history edge coverage, and is set to empty at the beginning (line 1). For all the initial seeds, \tool{} tests all of them and adds the seeds to the set $M$ if they discover new coverage (lines 2-6). Along with the seed, the set $M$ also includes the number of new edges discovered by the seed. 
Then, \tool{} merges the current bitmap $B$ into the overall bitmap $B_{overall}$ (line 7). 
Based on the rankings of seeds, \tool{} selects a top seed from the seed corpus (line 10). 
After assigning the energy (\ie the number of mutations) to the selected seed (line 12), \tool{} mutates the seed based on the \textit{fitness} computed in Byte Analysis phase and generates a new input (line 13). After the program is executed with the new input, fuzzing checks whether the corresponding bitmap has at least a non-zero element that has never been covered before (lines 14-15). If so, \tool{} counts the number of such non-zero elements, and retains the new input as a new seed (lines 16-17). 
Then, the energy decreases by one and the current bitmap is merged into the overall bitmap (lines 18-19).
According to the number of new edges discovered by mutating the seed $s$, $n$ will be updated with $n_{all}$.
\tool{} sorts the seeds in the seed corpus based on the number of new edges found (lines 22).

\subsection{Application}
Our approach can be applied to other CGFs without much efforts.
The improvement is added to the stages when fuzzing selects a seed and when fuzzing selects a byte. The seed corpus is ordered based on seed rankings so that other fuzzers select seeds from the top of the ordered sequence. 
Most CGFs randomly select bytes to be mutated or select bytes that are related to certain path constraints. After other CGFs select a byte, our \tool{} will accept or reject the selection based on the probability distributions of bytes. In order to obtain the probability distributions, \tool{} integrates our Byte Analysis into the deterministic stage if other fuzzers have the stage. This stage mutates seeds byte by byte based on the determined mutation strategy. If a fuzzer does not perform the deterministic stage, \tool{} conducts a pre-mutation to obtain probabilities.

\section{Evaluation}
In this section, we run experiments to verify the performance of \tool{} and to answer the following research questions: 

\begin{itemize}[leftmargin=*]
    \item \textbf{RQ1:} How effective is Byte Analysis to guide more generated inputs flow into functional code? 
    \item \textbf{RQ2:} To what extent can avoiding mutating validation-related bytes and how can Seed Prioritization improve the efficiency of fuzz testings? 
    \item \textbf{RQ3:} How many bugs can be (uniquely) discovered by the proposed framework? 
\end{itemize}

\begin{table}[!t]
\centering
\caption{Target programs.}
\resizebox{0.9\linewidth}{!}{
\begin{threeparttable}
\begin{tabular}{lccl}
    \toprule
    \textbf{Target} & \textbf{Source file} & \textbf{Input format} & \textbf{Test instruction} \\ \midrule
    \textbf{tiff2bw} & \multirow{2}{*}{libtiff-4.0.9} & \multirow{2}{*}{tiff} & tiff2bw @@\tnote{1}~{ } /dev/null \\
    \textbf{tiffsplit} &&& tiffsplit @@ /dev/null\\
    \midrule
    \textbf{nm} & \multirow{5}{*}{binutils-2.30} & \multirow{5}{*}{ELF} & nm -C @@ \\
    \textbf{objdump} &&& objdump -D @@ \\
    \textbf{readelf} &&& readelf -a @@ \\
    \textbf{size} &&& size @@ \\
    \textbf{strip} &&& strip -o tmp\_file @@  \\
    \midrule
    \textbf{djpeg} & libjpeg-9c & jpeg & djpeg @@ \\
    \bottomrule
\end{tabular}
\begin{tablenotes}
\item[1] @@: A placeholder indicating that the input is a file.
\end{tablenotes}
\end{threeparttable}

}
\label{subjects}
\end{table}

\begin{table*}[!t]
\centering
\caption{Original vs. \textsc{Truzz}-equipped fuzzers on the percentage of valid input generated.} \label{tab:valid_percent}
\resizebox{0.9\linewidth}{!}{
\begin{threeparttable}
\begin{tabular}{@{\extracolsep{4pt}}lcrrrrrrrrr@{}}
\toprule
\multirow{3}{*}{\textbf{Program}} & \multirow{3}{*}{\makecell[c]{\textbf{\# of validation}\\\textbf{checks}}} & \multicolumn{4}{c}{\textbf{NEUZZ}} & \multicolumn{4}{c}{\textbf{\tool{} (NEUZZ)}\tnote{1}} & \multirow{3}{*}{\begin{tabular}[c]{@{}c@{}}\textbf{Improvement of}\\ \textbf{Valid Ratio}\end{tabular}}\\
\cline{3-6} \cline{7-10}

& &\multicolumn{3}{c}{\textbf{\# of inputs generated}} & \multirow{2}{*}{\begin{tabular}[c]{@{}c@{}}\textbf{Valid}\\ \textbf{Ratio}\end{tabular}} & \multicolumn{3}{c}{\textbf{\# of inputs generated}} & \multirow{2}{*}{\begin{tabular}[c]{@{}c@{}}\textbf{Valid}\\ \textbf{Ratio}\end{tabular}} &\\
\cline{3-5} \cline{7-9}

&& \textbf{Valid}\tnote{2} & \textbf{Invalid}\tnote{3} & \textbf{Total} &  & \textbf{Valid} & \textbf{Invalid} & \textbf{Total} & &\\
\midrule
\textbf{nm} & 41 & 266,119 & 6,689,897 & 6,956,016 & 3.82\% & 2,531,141 & 4,424,875 & 6,956,016 & 36.38\% & 32.56\%\up  \\
\textbf{objdump} & 46 & 537,032 & 6,566,985 & 7,104,017  & 7.55\% & 1,957,187 & 5,146,830 & 7,104,017 & 27.55\% & 20.00\%\up \\
\textbf{readelf} & 183 & 528,506 & 6,645,851 & 7,174,357 & 7.36\% & 866,535 & 6,307,822 & 7,174,357 & 12.07\% & 4.71\%\up \\
\textbf{size} & 11 & 543,830 & 6,396,120 & 6,939,950 & 7.83\% & 1,576,477 & 5,363,473 & 6,939,950  & 22.71\% & 14.88\%\up \\
\textbf{strip} & 150 & 460,160 & 6,783,308 & 7,243,468 & 6.35\% & 1,079,114 & 6,164,354 & 7,243,468  &14.90\% & 8.55\%\up \\
\bottomrule
\end{tabular}
\begin{tablenotes}
\item[1] \textbf{\tool{} (NEUZZ)}: an implementation of \tool{} framework using NEUZZ as the fuzzer.
\item[2] \textbf{Valid}: indicates that the inputs pass all validation checks.
\item[3] \textbf{Invalid}: indicates that the inputs flow into error-handling code.
\end{tablenotes}
\end{threeparttable}
}
\end{table*}

\subsection{Experimental Setup}
In order to evaluate our \tool, we implemented \tool{} with 6 state-of-the-art fuzzers and compared the performance with the corresponding vanilla fuzzers.
Specifically, we ran each fuzzer on 8 target programs
for 24 hours, and then repeated the experiment for 5 times. All our measurements were performed on a system running Ubuntu 18.04 with Intel(R) Xeon(R) Gold 6230R CPU and 4 NVIDIA GeForce RTX 2080 Ti GPUs.

\noindent \textbf{Benchmark Fuzzers.} As mentioned earlier, we select 6 fuzzers to evaluate our approach. NEUZZ~\cite{neuzz} and GreyOne~\cite{GreyOne} are selected because they infer byte-constraint relation to resolve constraints. We choose AFL~\cite{afl} because it is a general CGF.
We also choose some extensions of AFL.
AFLFast~\cite{aflfast} improves AFL via formulating path transitions. MOPT~\cite{MOPT} optimizes AFL's mutation operator schedule. FuzzFactory~\cite{fuzzfactory} generalizes coverage-guided fuzzing to domain-specific testing goals.

\begin{itemize}[leftmargin=*]
    \item \textbf{NEUZZ}  models relations between input bytes and branch behaviors by utilizing neural networks. 
    In order to apply Byte Analysis to NEUZZ, \tool{}~(NEUZZ) performs a pre-mutation before the mutation strategy proposed by NEUZZ. 
    \item \textbf{GreyOneFTI}  utilizes a lightweight and sound fuzzing-driven taint inference (FTI) to infer the relations between input bytes and variables in path constraints. We name the fuzzer as GreyOneFTI because GreyOne does not release its code and we try our best efforts to implement its core of FTI. 
    \tool~(GreyOneFTI) improves GreyOneFTI via protecting the bytes related to the outer conditions. 
    
    \item \textbf{AFL}, the American Fuzzy Lop, is a widely-used CGF, which has discovered many bugs in real-world applications. 
    \tool(AFL) infers the validation-related bytes during its deterministic stage, and selects bytes based on the byte probability distribution during its havoc stage. 
    \item \textbf{AFLFast} is developed based on AFL and optimizes power schedule via formulating path transitions. As a result, AFLFast prefers assigning more computing resources to the paths that are exercised with low frequency. Similar to \tool~(AFL), \tool~(AFLFast) applies byte probability to select bytes.  
    \item \textbf{MOPT} utilizes Particle Swarm Optimization (PSO) algorithm to achieve mutation operator schedule. 
    \tool~(MOPT) applies Byte Analysis in all types of fuzzing modes to protect the bytes related to validation checks. we set the parameter \textit{-L} as 60. MOPT provides two types of pacemaker fuzzing modes and we choose \textit{MOPT-AFL-tmp} for the experiments. 
    \item \textbf{FuzzFactory} ~\cite{fuzzfactory} saves intermediate inputs if they make progress in domain-specific states.  The implementation of \tool~(FuzzFactory) is similar to that of \tool~(AFL).
\end{itemize}

\noindent \textbf{Target Programs.} We evaluated \tool{} on 8 different real-world programs (as shown in Table~\ref{subjects}), which we adopted from the evaluation of NEUZZ~\cite{neuzz} and MOPT~\cite{MOPT}. 
\texttt{nm} lists the symbols from object files. \texttt{objdump} displays information about object files. \texttt{readelf} displays information about ELF format object files. \texttt{size} lists the section sizes for each of the binary files. \texttt{strip} discards all symbols from object files. \texttt{djpeg} is a widely-used tool for handling JPEG image files. \texttt{tiff2bw} converts a color TIFF image to greyscale. \texttt{tiffsplit} splits a multi-image TIFF into single-image TIFF files.

\noindent \textbf{Initial Seeds.} To make the comparison fair, we run each fuzzer on each target application with the same initial seeds collected from target program's test suite and public seed corpus(NEUZZ\cite{neuzz} and MOPT\cite{MOPT}).
NEUZZ is different from other fuzzers because it requires an initial dataset. 
According to the setup described in NEUZZ~\cite{neuzz}, we run AFL for an hour to collect an initial training set $(X, Y)$, where $X$ is a set of input bytes and $Y$ represents the corresponding edge coverage bitmap. 
The average size of 8 initial training set is 1,005.

\begin{figure*}[!t]
\centering
\includegraphics[width=1\linewidth]{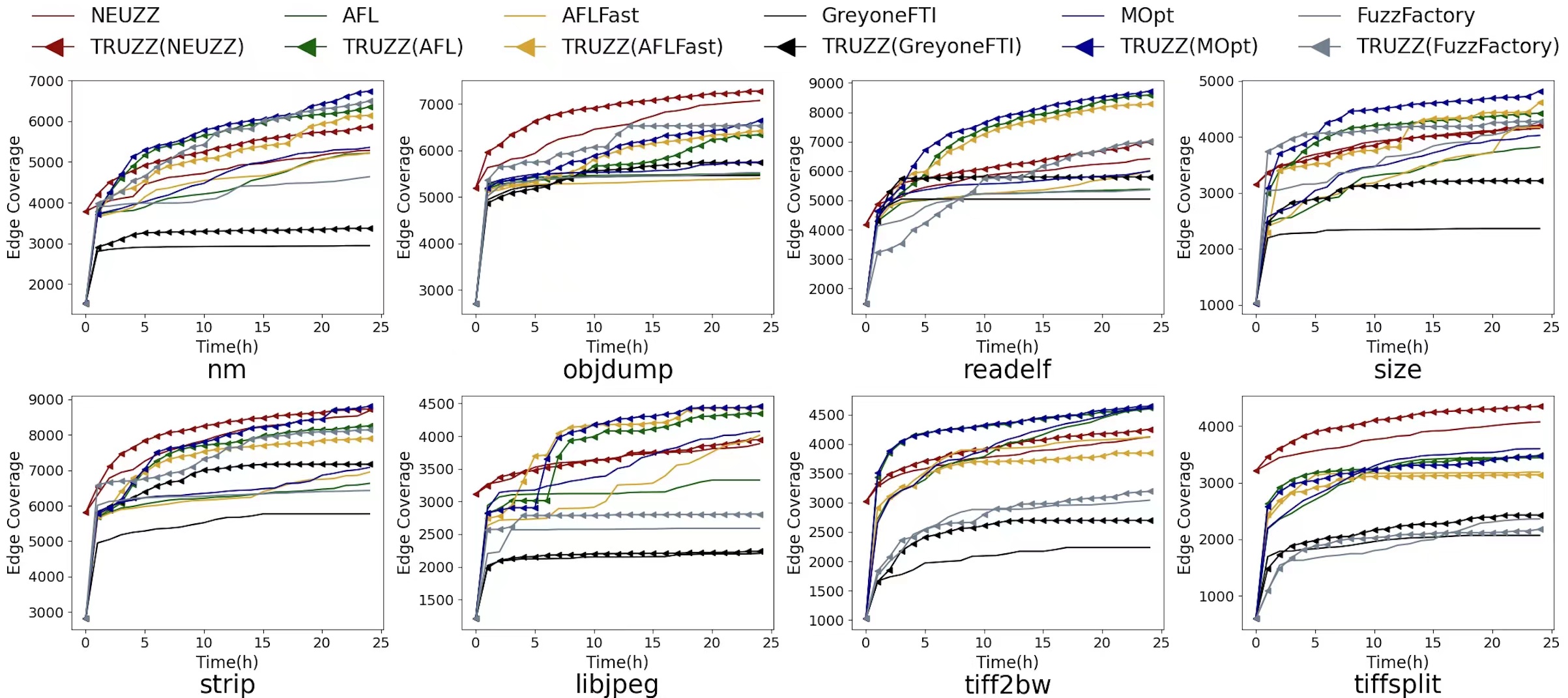}
\caption{The edge coverage of different fuzzers running for 24 hours.}
\label{fig:edge_time}
\end{figure*}

\subsection{RQ1: Effectiveness of Byte Analysis} \label{sec:eval_byte}
As described in the methodology, \tool{} uses Byte Analysis to identify bytes associated with validation checks.
Then, we will mutate the validation-related bytes much less often than other bytes.
Therefore, \tool{} relies heavily on the accuracy of Byte Analysis.
As such, this subsection analyzes such accuracy of \tool's Byte Analysis.
We manually analyze five programs utilized in the experiment to mark all error-handling paths in these programs, including \texttt{nm}, \texttt{objdump}, \texttt{readelf}, \texttt{size}, and \texttt{strip}.
For the analysis methods, we adopt the manual analysis method used by FIFUZZ~\cite{FIFUZZ}.
We scan the definition of each error function, and check whether it can trigger an error by returning an error code or a null pointer.
Each time a mutated input is generated and executed, we determine whether the input exercises the error-handling path and count the number of those paths during fuzzing. An input is considered ``valid'' if it passes all validation checks and triggers no error-handling code.

To avoid the impact of Seed Prioritization, we implement the \tool{} only with Byte Analysis in this experiment. 
Finally, we take NEUZZ and \tool{}(NEUZZ) as an example to illustrate the accuracy of Byte Analysis.
In NEUZZ's mutation strategy, it randomly selects 500 edges and takes the derivative of two seeds for each edge. Then the gradient information obtained by derivation is recorded in the gradient file. Finally, NEUZZ will mutate based on this gradient file.
In order to ensure the fairness of the experiment, 1,000 seeds are randomly selected and the same model is used for both fuzzers.
Both NEUZZ and \tool{}(NEUZZ) mutate inputs based on the same gradient file, and both of them have the same number of mutations.
We calculate the proportion of valid inputs in their generated mutated inputs, as shown in Table~\ref{tab:valid_percent}. \tool{}(NEUZZ) generates on average 16.14\% more  valid samples than NEUZZ. Especially for \texttt{nm} and \texttt{objdump}, \tool{}(NEUZZ) generates over 20\% more valid samples than NEUZZ.

\vspace{1mm}
\begin{mdframed}[backgroundcolor=white!10,rightline=true,leftline=true,topline=true,bottomline=true,roundcorner=2mm,everyline=true] 
	\textbf{Answer to RQ1.} Under the same experimental conditions, \tool{} can generate more valid samples than the vanilla fuzzers. 
\end{mdframed}

\begin{table*}[!t]
\centering
\caption{Original vs. \textsc{Truzz}-equipped fuzzers on the number of new edges found in 24 hours.}
\resizebox{\linewidth}{!}{
\begin{threeparttable}
\begin{tabular}{@{\extracolsep{4pt}}lrrrrrrrrrrrr@{}}
\toprule
\multirow{3}{*}{\textbf{Program}} & \multicolumn{12}{c}{\textbf{\# of New Edges}}\\
\cline{2-13}
& \multirow{2}{*}{\textbf{AFL}} & \textbf{\tool} & \multirow{2}{*}{\textbf{AFLFast}} & \textbf{\tool} & \multirow{2}{*}{\textbf{NEUZZ}} & \textbf{\tool} & \multirow{2}{*}{\textbf{MOPT}} & \textbf{\tool} & \multirow{2}{*}{\textbf{FuzzFactory}} & \textbf{\tool} & \multirow{2}{*}{\textbf{GreyoneFTI}} & \textbf{\tool}\\
&&  (AFL) & &  (AFLFast) &  &  (NEUZZ) &  &  (MOPT) &  &  (FuzzFactory) &  &  (GreyoneFTI) \\
\midrule

\textbf{nm} & 3,690 & 4,828(30.84\%\up) & 3,694 & 4,620(25.07\%\up) & 1,477 & 1,989(34.66\%\up) & 3,829 & 5,219(36.30\%\up) & 3,109 & 4,970(59.86\%\up) & 1,417  & 1,845(30.20\%\up)\\
\textbf{objdump} & 2,779 & 3,622(30.33\%\up) & 2,682 & 3,707(38.22\%\up) & 1,875 & 2,092(11.57\%\up) & 3,041 & 3,929(29.20\%\up) & 2,808 & 3,822(36.11\%\up) & 2,747 & 3,025(10.12\%\up) \\
\textbf{readelf} & 3,906 & 7,147(82.97\%\up) & 4,521 & 6,778(49.92\%\up) & 2,166 & 2,707(24.98\%\up) & 4,513 & 7,231(60.23\%\up) & 3,866 & 5,522(42.83\%\up) & 3,545  & 4,302(21.35\%\up)\\
\textbf{size} & 2,789 & 3,394(21.69\%\up) & 3,142 & 3,583(14.04\%\up) & 985 & 1,048(6.40\%\up) & 2,994 & 3,793(26.69\%\up) & 3,239 & 3,245(0.19\%\up) & 1,333  & 2,186(63.99\%\up)\\
\textbf{strip} & 3,821 & 5,444(42.48\%\up) & 4,140 & 5,086(22.85\%\up) & 2,804 & 2,861(2.03\%\up) & 4,270 & 5,989(40.26\%\up) & 3,620 & 5,331(47.27\%\up)  & 2,961  & 4,360(47.25\%\up)\\
\textbf{libjpeg} & 2,122 & 3,145(48.21\%\up) & 2,808 & 3,240(15.38\%\up) & 755 & 844(11.79\%\up) & 2,862 & 3,254(13.70\%\up) & 1,384 & 1,602(15.75\%\up) & 1,000  & 1,031(3.10\%\up)\\
\textbf{tiff2bw} & 3,572 & 3,597(0.70\%\up) & 3,090 & 2,819(8.77\%\down) & 1,098 & 1,192(8.56\%\up) & 3,582 & 3,627(1.26\%\up) & 2,017 & 2,168(7.49\%\up) & 1,212  & 1,674(38.12\%\up) \\
\textbf{tiffsplit} & 2,844 & 2,886(1.48\%\up) & 2,592 & 2,545(1.81\%\down) & 845 & 1,130(33.73\%\up) & 3,005 & 2,869(-4.53\%\down) & 1,763 & 1,582(-10.27\%\down) & 1,473  & 1,833(24.44\%\up)\\
\bottomrule
\end{tabular}
\end{threeparttable}

}
\label{tab:edge_coverage}
\end{table*}

\begin{table}[!t]
\centering
\caption{The number of new edges found compared to vanilla fuzzers in 12 hours.}
\resizebox{0.95\linewidth}{!}{
\begin{threeparttable}
\begin{tabular}{@{\extracolsep{4pt}}lrrrrrrrrrrrr@{}}
\toprule
\textbf{Fuzzer} & \textbf{Average} & \textbf{Min.} & \textbf{Max.} & \textbf{Median} \\ 
\midrule
\textbf{BA-equipped\tnote{1}\;  NEUZZ} & +14.84\% & +8.06\% & +27.34\% & +12.79\% \\ 
\textbf{SP-equipped\tnote{2}\;  NEUZZ} & +11.51\% & -2.97\% & +33.05\% & +8.63\% \\ 
\textbf{BA-equipped GreyOne} & +17.43\% & +1.72\% & +42.73\% & +17.96\% \\ 
\textbf{SP-equipped GreyOne} & +52.16\% & +13.91\% & +147.45\% & +42.07\% \\ 
\textbf{BA-equipped AFL} & +0.58\% & -0.77\% & +1.65\% & +0.71\% \\ 
\textbf{SP-equipped AFL} & +30.83\% & -3.95\% & +55.56\% & +40.44\% \\ 
\textbf{BA-equipped AFLFast} & -0.39\% & -4.06\% & +6.07\% & -0.7\% \\ 
\textbf{SP-equipped AFLFast} & +16.90\% & -16.51\% & +44.42\% & +18.29\% \\ 
\textbf{BA-equipped MOPT} & -1.20\% & -2.75\% & +1.67\% & -1.58\% \\ 
\textbf{SP-equipped MOPT} & +27.31\% & +1.03\% & +49.39\% & +31.47\% \\ 
\textbf{BA-equipped FuzzFactory} & -1.84\% & -4.50\% & +0.10\% & -1.82\% \\ 
\textbf{SP-equipped FuzzFactory} & +33.42\% & +16.69\% & +66.91\% & +26.74\% \\

\bottomrule
\end{tabular}
\begin{tablenotes}
\item[1] \textbf{BA-equipped}: Use Byte Analysis to optimize the vanilla fuzzer only.
\item[2] \textbf{SP-equipped}: Use Seed Prioritization to optimize the vanilla fuzzer only.

\end{tablenotes}
\end{threeparttable}
}
\label{tab:edge_coverage_separate}
\end{table}

\begin{table}[!t]
\centering
\caption{Results of Vargna and Delaney's $\hat{A}_{12}$ scoring.}
\resizebox{\linewidth}{!}{
\begin{tabular}{@{\extracolsep{4pt}}lrrrrrrrr@{}}
\toprule

\textbf{Program} &  \textbf{\tool} & \textbf{\tool}& \textbf{\tool} & \textbf{\tool} & \textbf{\tool} & \textbf{\tool}\\
&  \textbf{(AFL)} & \textbf{(AFLFast)} & \textbf{ (GreyOneFTI)} & \textbf{(NEUZZ)} &  \textbf{(MOPT)} & \textbf{(FuzzFactory)}\\
\midrule
\textbf{nm}  & 1.00  & 1.00  & 1.00  & 1.00 &1.00 &1.00 \\
\textbf{objdump}  & 1.00  & 1.00  & 1.00  & 1.00 &1.00 &1.00 \\
\textbf{readelf}  & 1.00  & 1.00  & 1.00  & 0.94 &1.00 &1.00 \\
\textbf{size}  & 1.00  & 1.00 & 1.00 & 0.81 &1.00 &0.33 \\
\textbf{strip} & 1.00 & 1.00 & 1.00 & 0.66 &1.00 &1.00\\
\textbf{libjpeg} & 1.00 & 0.94 & 1.00 & 0.69 &1.00 &1.00 \\
\textbf{tiff2bw} & 0.62 & 0.00 & 1.00 & 0.88 &0.58 &1.00 \\
\textbf{tiffsplit} & 0.73 & 0.42 & 1.00 & 1.00 &0.13 &0.00 \\

\bottomrule
\end{tabular}
}
\label{A12}
\end{table}

\subsection{RQ2: Code Coverage Discovery}\label{sec:code_coverage}

A bug is triggered only when the related code regions are explored. 
Therefore, code coverage is an important metric for coverage-guided fuzzers. 
We run two versions of fuzzers on eight programs for 24 hours (repeated five times) and leverage AFL's afl-showmap tool\cite{afl} to compute the code coverage. Then we compared the average performance with respect to edge coverage. We first evaluated the growth trend of edge coverage by two versions of fuzzers. As shown in Fig.~\ref{fig:edge_time}, \textsc{Truzz}-equipped fuzzers have a steady and stronger growth trend on all target programs except for \texttt{tiff2bw} and \texttt{tiffsplit}. Note that, for NEUZZ, as there are 1,000 initial seeds involved, the edge coverage at the starting point is much higher than other fuzzers with only one initial seed.

Specifically, \tool{}-equipped fuzzers achieve better new edges coverage, 
for most of the programs than the vanilla versions as shown in Table~\ref{tab:edge_coverage}. 
On average, \tool-equipped fuzzers discovered 24.75\% more new edge coverage than vanilla fuzzers.
For example, \tool{} (AFL) discovers more than 20\% edges than AFL in 6 target programs.
However, due to AFL and AFLFast do not infer the byte-constraint relation, the reason why the code coverage decreases in some cases is still not clear. To reveal a possible root cause, we provide a case study in Section~\ref{sec:real_threat}. It would be interesting to ascertain the root cause in the future work.

We further evaluate the performance of \textsc{Truzz} using Vargha-Delaney A Measurement~\cite{vargha2000critique}, where $\hat{A}_{12} \in [0,1]$ denotes the measure of stochastic superiority that population 1 is superior to population 2. If the two populations are equivalent, then $\hat{A}_{12} = 0.5$. 
An $\hat{A}_{12}$ larger than 0.5 indicates that the population 1 is \textit{stochastically larger} than the  population 2, and vice versa. For example, the difference between \textsc{Truzz}-equipped fuzzer and vanilla fuzzer is determined as (1) big when $\hat{A}_{12} \geq 0.71$; (2) medium when $\hat{A}_{12} \geq 0.64$; and (3) small when $\hat{A}_{12} \geq 0.56$.
Specifically, we test the numbers of new edges and path coverage collected from the five rounds of experiments by vanilla and \tool-equipped fuzzers after 24 hours.
As shown in Table~\ref{A12}, 
the percentage of the results that $\hat{A}_{12} \geq 0.71$ is 81.25\%. Especially, 70.83\% of the results achieve the highest score of $\hat{A}_{12} = 1.00$. Therefore, we conclude that the difference in new edge and path coverage between vanilla fuzzer and \textsc{Truzz}-equipped fuzzer is statistically large.

Further, We conducted separate experiments on Byte Analysis and Seed Prioritization in \textsc{Truzz}, as shown in Table \ref{tab:edge_coverage_separate}. Byte Analysis equipped fuzzers (NEUZZ and GreyOne) and Seed Prioritization equipped fuzzers (all fuzzers) discovered
16.13$\pm$10.25\% and 28.68$\pm$25.85\% more new edges than vanilla fuzzers, respectively.
The performance of Byte Analysis equipped AFL, AFLFast, MOPT, and FuzzFactory does not improve much. We believe the reason is that those fuzzers randomly select mutation positions, resulting in a low selection probability of mutating validation-related bytes.

Inference-based fuzzers such as NEUZZ and Greyone will firstly mutate interesting bytes(\eg the bytes related to edges). However, based on their inference strategy, validation-related bytes are often considered as "interesting". If validation-related bytes are not protected during the mutation phase, mutated inputs will flow into the error-handling code.
Therefore, Byte Analysis is more suitable for inference-based fuzzers, \ie, the fuzzers that infer the relationships between input bytes and path constraints.

\vspace{1mm}
\begin{mdframed}[backgroundcolor=white!10,rightline=true,leftline=true,topline=true,bottomline=true,roundcorner=2mm,everyline=true] 
	\textbf{Answer to RQ2.} \tool{} can significantly improve the coverage discovery (on average 24.75\% more edge coverage). In 44 out of 48 program-fuzzer pairs, \tool{}-equipped fuzzers achieve more code coverage than their vanilla fuzzers.
\end{mdframed}

\subsection{RQ3: Bug Discovery}
We compare \textsc{Truzz}-equipped fuzzers with vanilla fuzzers in terms of the number of real-world bugs.

We use GDB and afl-collect\cite{afl_collect} to analyze the number of bugs discovered by a fuzzer.
First, we use afl-collect to remove invalid crash samples and achieve crash sample de-duplication.
Second, we use GDB to manually analyze the program logic for each bug, removing bugs that have the same root cause.
Table~\ref{tab:bug_analysis} shows the number of unique bugs (accumulated in 5 runs) found in the eight real-world applications.

In total, vanilla fuzzers and \textsc{Truzz}-equipped fuzzers find 10 and 13 vulnerabilities in four applications, respectively. 
In the remaining four programs, all the fuzzers do not identify bugs in 24 hours.
Specially, \texttt{strip} is reported as vulnerable only by \textsc{Truzz}-equipped fuzzers. 
For \texttt{tiff2bw}, it converts an RGB or Palette color TIFF image to a greyscale image by combining percentages of the red, green, and blue channels. We discover 9 bugs in \texttt{tiff2bw}. 
In the case of \texttt{readelf}, we found one bug when it displays information from any ELF format object file.
\texttt{strip} displays a list showing all architectures and object formats available, and we identify 3 bugs.

For the three \textsc{Truzz}-missed bugs (two in \texttt{size} discovered by AFL and MOPT, and one in \texttt{tiff2bw} discovered by MOPT), we found that the execution speeds of MOPT (in \texttt{size} and \texttt{tiff2bw}) and AFL in \texttt{size} dropped 10\% and 13.55\%, respectively, which may lead to the missing of bugs. Further, the randomness of fuzzing may also contribute to the results. We repeated experiments 5 times for each fuzzer, but the \textsc{Truzz}-missed bugs were only discovered by AFL or MOPT in 1 trial.

\vspace{1mm}
\begin{mdframed}[backgroundcolor=white!10,rightline=true,leftline=true,topline=true,bottomline=true,roundcorner=2mm,everyline=true] 
	\textbf{Answer to RQ3.} \textsc{Truzz}-equipped fuzzers outperform their vanilla fuzzers in terms of bug discovery, identifying 13 bugs in 8 target programs and 6 of them cannot be identified by vanilla fuzzers.
\end{mdframed}
\vspace{-3mm}

\begin{table}[t!]
\centering
\caption{Bugs discovered by vanilla fuzzers (Orig.) and \textsc{Truzz}-equipped fuzzers (\tool(*)).}
\resizebox{0.75\linewidth}{!}{
\begin{tabular}{llcc}
\toprule
\multirow{2}{*}{\textbf{Program}} & \multirow{2}{*}{\textbf{Vulnerability Type}} & \multicolumn{2}{c}{\textbf{\# of Vulnerabilities}} \\
\cline{3-4}
& & \textbf{Orig.} & \textbf{\tool(*)} \\
\midrule
\textbf{readelf} & AbortSignal & 1 & 1\\
\hline
\multirow{2}{*}{\textbf{size}} & AbortSignal & 1 & 0\\
&  DestAvNearNull & 1  & 0\\
\hline
\multirow{3}{*}{\textbf{strip}} & AbortSignal & 0  & 1\\
&  DestAvNearNull & 0  & 1\\
&  SourceAv & 0  & 1\\
\hline
\multirow{5}{*}{\textbf{tiff2bw}} & AccessViolation & 2  & 1\\
& BlockMoveAv & 1  & 1\\
& BadInstruction & 0 & 1\\
& SourceAv & 2 & 3\\
& HeapError & 2 & 3\\

\bottomrule
\end{tabular}
}
\label{tab:bug_analysis}
\end{table}

\vspace{1mm}
\section{Discussion}

\textbf{Complexity of programs.} Our \tool{} infers the relations between input bytes and validation checks, but the inference may erroneously recognize non-validation checks as validation checks. That is, \tool{} may protect bytes that are related to non-validation checks. However, our \tool{} still allows fuzzing to examine the ``shorter'' path with a low probability, \ie the code regions belonging to the ``shorter'' path are still examined. One possible improvement is to optimally switch between the vanilla strategy and our Byte Analysis for selecting bytes. Another threat to \tool{} is that some longer paths may have smaller number of descendants. Thus, the priorities of some seeds may be erroneously assigned. \tool{} mitigates this problem via decreasing the rankings of seeds if mutations of them do not identify more new edges.

\noindent\textbf{Overhead of execution speed.}
\textsc{Truzz} will decrease the execution speed because it leads fuzzing to focus more on the execution of functionality code, which usually cost more computational time than error-handling code. 
The execution speed of Byte Analysis equipped GreyOne and NEUZZ dropped by 23.55±30.86\% and 2.50±1.94\%, separately. The execution speed of Seed Prioritization equipped AFL and AFLFast dropped by 12.22\% and 22.52\%, respectively. Specifically, the execution speed of AFLFast decreased significantly on \texttt{tiff2bw} and \texttt{tiffsplit} (59.74\% and 38.47\%, respectively). This could be the reason why \textsc{Truzz}(AFLFast) discover less new coverage.
We believe that such a trade-off is worthy because focusing on exploring functionality code will lead to more new code coverage and bugs finding.

\noindent\textbf{Limitations of \tool{}.} \textsc{Truzz} has limitations in handling the scenarios that targeting input bytes that may take on a variety of constants. For example, if the \textit{default} branch of \textit{switch-case} branches is an error-handling code, our \tool{} cannot distinguish between different \textit{cases}. The situation can be mitigated by setting a higher $L_p$ in Equation (\ref{equ:prob}). Another way to mitigate the situation is to involve more seeds, which may cover different branches in a \textit{switch-case}. Our \tool{} also has limitations when inferring bytes for checksum. A slight change of checksum-related data will fail the check of checksum, and \tool{} will avoid mutating checksum-related data. This hinders the exploration of different data values for checksum. As a future work, it is interesting to investigate on how to protect a block of contiguous bytes and how to actively solve constraints.

\section{Conclusion}
The inference approaches may include extra bytes that belong to validation checks. This will reduce the performance of fuzzing and assign too many computing resources to error-handling code. We mitigate the problem by inferring the validation-related bytes and protect the bytes from being frequently mutated. The inference is based on the observation that the difference of paths in path transitions imply the properties of programs. As a result, most of the generated inputs will examine the functional code, in lieu of error-handling code. This solution increases the probability of identifying more code coverage and bugs in functional code.
We design and implement \tool{} based on our idea, which increases the probability of generating valid inputs, \ie the inputs that flow into functional code only. 
Our evaluation on eight open-source programs shows that \textsc{Truzz}-equipped fuzzers significantly outperform their vanilla versions. We hope the optimization developed in this paper could advance the coverage-guided greybox fuzzing domain.

\section*{Data Availability}
To enable reproducibility, we provide a replication package publicly available at \url{https://github.com/truzz-fuzz/truzz-fuzz}.

\begin{acks}
This work was supported in part by the National Natural Science Foundation of China (61972219), the Research and Development Program of Shenzhen (JCYJ20190813174403598, SGDX201909181012016\\96), the National Key Research and Development Program of China (2018YFB1800601), and the Overseas Research Cooperation Fund of Tsinghua Shenzhen International Graduate School (HW2021013).
\end{acks}

\balance
\bibliographystyle{ACM-Reference-Format}
\bibliography{main-final}


\end{document}